\documentclass[showpacs,aps,pre,superscriptaddress,floatfix]{revtex4}
\usepackage{graphicx} 
\usepackage{mathtools}
\usepackage{amsmath}

\begin{document}

\title{Roughening of $k$-mer growing interfaces in stationary regimes}

\author{M. D. Grynberg} 

\affiliation{IFLP-CONICET, Departamento de F\'isica, Universidad Nacional de La Plata, 
1900 La Plata, Argentina}

\author{F. I. Schaposnik Massolo} 

\affiliation{Center for Theoretical Physics of the Universe, Institute for Basic Science
(IBS), Seoul 08826, Republic of Korea}

\begin{abstract}
We discuss the steady state dynamics of interfaces with periodic boundary 
conditions arising from body-centered solid-on-solid growth models in $1+1$ 
dimensions involving random aggregation of extended particles (dimers, 
trimers,\,$\cdots,k$-mers). Roughening exponents as well as width and 
maximal height distributions can be evaluated directly in stationary regimes 
by mapping the dynamics onto an asymmetric simple exclusion process with 
$k$-\,type of vacancies. Although for $k \ge 2$ the dynamics is partitioned into 
an exponentially large number of sectors of motion, the results obtained in 
some generic cases strongly suggest a universal scaling behavior closely 
following that of monomer interfaces.
\end{abstract}


\maketitle

\vskip -0.07cm
Because of its ubiquity in nature and importance in technology, the dynamics of growing 
interfaces has been investigated extensively for more than three decades in a vast body 
of experimental, theoretical, and numerical works \cite{Meakin, Odor}. Despite the 
diversity of morphologies in which growing interfaces can evolve, most of those studies 
pointed out the onset of scaling regimes emerging at both large time and length scales. 
This enabled a classification of seemingly dissimilar processes in terms of universality 
classes characterized by a set of scaling exponents which take over the late evolution 
stages \cite{Odor,Henkel}. It is by now well established that many discrete 
nonequilibrium growth models in one dimension (1D) evolving under a variety of simple 
stochastic rules belong to the Kardar-Parisi-Zhang (KPZ) universality class \cite{Meakin, 
Odor, Henkel, KPZ}. This latter effectively captures the statistical fluctuations of a set of 
heights $h_1(t),\,\cdots,\, h_L(t)$  growing at $L$ locations of a 1D substrate at a given 
time $t$. Starting from an initially flat substrate, the roughness or width developed by 
such discrete interfaces is often studied in terms of their mean square height fluctuations 
which, on general grounds, can be expected to follow the Family-Vicsek dynamic scaling 
ansatz \cite{Family}
\begin{equation}
\label{width}
\langle\, W^2 (L,t) \,\rangle = \frac{1}{L} \,\sum_n \left\langle \left[\, h_n (t)\, - \,\bar h(t) 
\,\right]^2\right\rangle \simeq L^{2 \zeta}\, f \left( t / L^z\right)\,,
\end{equation}
for large substrate sizes. Here $\bar h(t)$ is the average height at instant $t$ of a given 
configuration (in turn being averaged by the outer brackets), whereas $f(x)$ refers to a 
universal scaling function behaving as $x^{\zeta/z}$ for $x \ll 1$, while approaching a 
constant for $x \gg 1$. Thus, at early stages the width is expected to grow as 
$t^{\zeta/z}$ until saturating as $L^{2\zeta}$ for times larger than $L^z$. The dynamic 
exponent $z$ therefore gives the fundamental scaling between length and time, whereas 
the Hurst or roughening exponent $\zeta$ measures the stationary dependence of 
$\langle\,W^2 \rangle$ on the typical substrate size. 

When it comes to this latter stationary aspect, note that the height levels of the interface 
can also be thought of as the visited sites of a 1D Brownian path extended on a time 
interval, here playing the role of the substrate length. Therefore, the usual root mean 
square displacement of normal random walks should constrain $\sqrt{\langle\, W^2
\rangle}$ to saturate as $L^{1/2}$, thus leaving us with a roughening exponent $\zeta
 = 1/2$. In fact this holds for numerous models of discrete interfaces, and is typical of 
both 1D KPZ and Edwards-Wilkinson (EW) \cite{Edwards} universality classes. However, 
in cases in which the path of the interface actually corresponds to a {\it correlated}
random walk, the stationary width may well saturate with subdiffusive exponents $\zeta 
< 1/2$. This anomalous scaling has been studied in even visiting random walks 
\cite{Nijs}, self-flattening and self-expanding interfaces \cite{Park}, as well as in the 
context of parity conserving growth processes \cite{Odor, Odor2, Arlego}. In particular, 
these latter involve the aggregation of composite objects \cite{note1}, which ultimately
causes the phase space to decompose into an exponential number of sectors of motion 
\cite{Arlego, Barma}. In this work we further consider the stationary dynamics of 
extended particles depositing over more than one height location at a time but where, 
despite the correlated walks associated to the paths of the interface, the usual diffusive 
width is restored. Moreover, as we shall see, our results also closely follow the entire 
width probability distribution of random walk interfaces \cite{Zia}, as well as the
distribution of their maximal heights measured with respect to the spatial average height 
already established both analytically and numerically for a wide set of solid-on-solid 
interfaces \cite{Shapir, Majumdar, Schehr}.

\begin{figure}[htbp]
\includegraphics[width=0.73\textwidth]{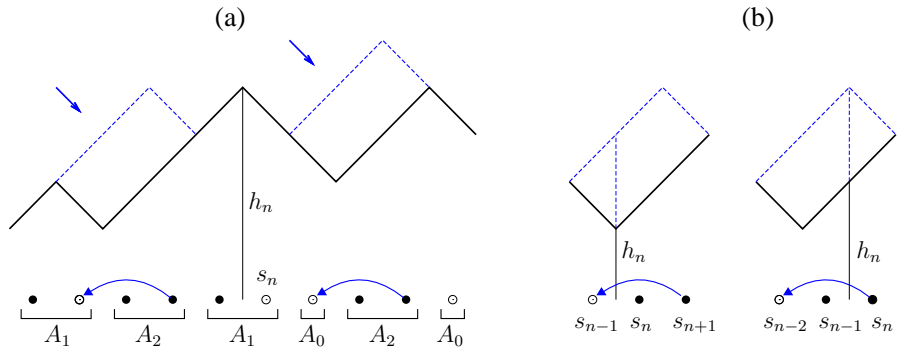}
\vskip -0.25cm
\caption{(a) Schematic view of a dimer-growing BCSOS interface (formed by slopes $s_n 
\equiv h_{n+1} - h_n = \pm 1$) and its equivalent driven lattice gas of reconstituting 
particles. At each step of the former, a dimer aggregation on local minima corresponds
in the latter to an exchange of $A_2$-\,`particles' with either of the $A_0$ or $A_1$  
`vacancies' referred to in the text. Note that under $A_1 A_2 \to A_2 A_1$ exchanges the 
indentity of $A_2$-\,dimers is not preserved. (b) Sublattice currents contributing to the 
growth velocity of $h_n$ as discussed through Eqs.\,\eqref{currents}-\eqref{velocity}. 
\vspace{-0.3cm}}
\label{processes}
\end{figure}

The process considered is a simple yet nontrivial extension of monomer adsorption in 
body-centered solid-on-solid (BCSOS) growth models \cite{Meakin,Plischke} whereby 
height differences $h_{n+1} - h_n$ between adjacent locations are restricted to 
$\pm 1$. Our basic kinetic steps involve the oblique incidence of extended particles, 
such as dimers, trimers,\,$\cdots,k$-mers, on the local minima of a BCSOS interface 
with periodic boundary conditions (PBC). An illustration of these processes is shown 
in Fig.\,\ref{processes} for the case of dimers. 
At each successful step $k$-contiguous locations increase their heights in two unit 
lengths, the rates of deposition being uniform and set equal to one per unit time. 
Thus, we see that the distance between a minimum and its nearest right maximum is 
preserved modulo $k$, in turn bringing about correlated movements and many-sector
decomposition of the interface walks.

The partitioning of the phase space of these paths can be understood with the aid of 
a mapping into a modification of the asymmetric simple exclusion process (ASEP) 
\cite{Ligget}, hereafter referred to as $k$-ASEP \cite{Barma2}. It consists of driven 
hard-core extended particles occupying $k$ consecutive sites while moving leftward by 
one site (e.g. $\circ \bullet \bullet  \rightarrow \bullet \bullet \circ$ say for dimers). Now, 
following Ref.\,\cite{Plischke}, if we think of these 0,1 occupancies  as stemming from 
Ising variables associated with the slopes $s_n = h_{n+1} - h_n$ of the interface (cf. 
Fig.\,\ref{processes}), it is then clear that  up to an immaterial constant its heights are 
obtained as $h_n = \sum_{j =1}^n s_j$. On the other hand since the interface is grown 
only out of $k$-mers, note that in the $k$-ASEP representation neither monomers nor 
groups or fragments of $j$-adjacent particles can move explicitly if $j < k$, although 
they are allowed to in a series of steps. For instance, in the sequence
\begin{equation}
\label{reconstruction}
0\, \underbracket[0.5pt]{1.\,.1}_j\, 0 \,\underbracket[0.5pt]{1 \dots 1}_k\; 
\rightarrow \;0\,  \underbracket[0.5pt]{1 \dots 1}_j \,
\underbracket[0.5pt]{1 .\,. 1}_k\,0 \; \rightarrow \; 
\underbracket[0.5pt]{1 \dots 1}_k\, 0 \,\underbracket[0.5pt]{1.\,.1}_j \,0 \,,
\end{equation}
the initial leftmost group of $j$ particles can hop $k$-sites to the right provided that
$k$-mers can dissociate and reconstitute, so they do not maintain their indentity 
throughout (except in the absence fragments; see below). In turn following 
Ref.\,\cite{Barma3}, these processes can also be interpreted as $k$-mer `particles'
$A_k $ moving through a set of $k$ composite characters or `vacancies' constructed as
\vskip -0.65cm
\begin{eqnarray}
\nonumber
A_0 & \equiv &  0\,,\\
\label{characters}
A_1 & \equiv &_{_{\hskip -0.39cm \vdots}}\;\;\,1\, 0\,,\\
\nonumber
A_j & \equiv & \underbracket[0.5pt]{1 \dots 1}_j\, 0\;, \;\; 1 <  j <  k\,.
\end{eqnarray}
The movements and recompositions of $k$-mers can then be thought of as character 
exchanges of the form $A_j \, A_k  \rightarrow A_k \, A_j$, the $k$-mer identity being 
preserved only by $A_0$, whereas exchanges not involving $A_k$ remain disabled, 
(\emph{i.e.} $A_i \, A_j $ do not swap their positions if  $i,j \ne k\,$). In this notation, for 
example the steps referred to in Eq.\,\eqref{reconstruction} now become $A_0\,A_j\,A_k 
\rightarrow A_0\,A_k\,A_j \rightarrow A_k\,A_0\,A_j$. But the key issue to point out here is 
that the $A_0, A_1,\cdots, A_{k-1}$ characters define a sequence or irreducible string (IS) 
whose ordering (set by the initial conditions) is {\it conserved} throughout all 
subsequent times. Thus, the invariant IS of a given sector of motion just refers to the 
succession of vacancy types obtained after deleting all $k$-mers or `reducible' characters 
appearing in any configuration of that sector. In other words, all states linked by the 
$k$-ASEP dynamics have the same IS.

{\it Effective ASEP}.--- 
Before evaluating the number of conservation laws yielded by this nonlocal construction, 
let us first remark that any state of these driven and reconstituting gases can be mapped 
to an  equivalent ASEP configuration defined on a smaller effective lattice \cite{Barma3}. 
More specifically, denoting by $N_j$ the number of $A_j$ characters (preserved 
throughout), and therefore given an IS sector of length ${\cal L} = \sum_{j \ne k} (j+1) 
N_j $, it is then clear that the $k$-ASEP dynamics amounts to an ASEP one with $N_k = 
(L- {\cal L})/k$ hard-core particles \cite{note2} driven through $N = \sum_j N_j$ sites; 
the effective density of such particles then being
\vskip -0.6cm
\begin{equation}
\label{density}
\rho_{_{ASEP}}^{-1} = 1+ \frac{k}{L - \cal L}\,\sum_{j \ne k} N_j\,.
\end{equation}
Thus, tagging the vacancies of a generic ASEP configuration in the same order as that 
appearing for the irreducible characters of a particular sector of motion, one can readily 
find the corresponding $k$-ASEP state just replacing the $n$-th ASEP vacancy by the 
$n$-th IS character, whilst substituting every ASEP particle in between by 
$k$-consecutive occupied sites. For instance, in an IS sector beginning as $A_1, A_0, A_0, 
A_1,\cdots$, say for dimers, the ASEP configuration $1\,0\,1\,0\,0\,1\,0\,\cdots$ will be 
mapped to $(11)(10)(11)(0)(0)(11)(10) \cdots$ $k$-ASEP occupancies. Now, recalling 
that under PBC the ASEP has a uniform steady state measure \cite{Ligget} (\emph{i.e.} 
all configurations are equally weighted), evidently it follows that this mapping will enable 
us to sample the steady state of generic IS sectors without explicitly evolving the 
$k$-ASEP in time. 

{\it Growth rates}.--- 
Under PBC the effective ASEP also allows for the evaluation of growth velocities. Since the 
original chain can be partitioned into $k$-sublattices $\Lambda_1, \cdots, \Lambda_k$ 
\cite{note2}, each $k$-mer covers one of their $L/k$ locations and so the $k$-ASEP 
dynamics preserves the monomer density per sublattice (also determined by the initial 
conditions). In turn, this defines $k$-stationary sublattice currents (eventually equivalent 
depending on the IS considered), given by
\begin{equation}
\label{currents}
J_{\alpha} = \big\langle \left(1 - n_{i-k}\right)\; n_{i-k+1} \cdots\,n_i\, \big\rangle\,,\; 
\forall \, i \in \Lambda_{\alpha}\;,\; \alpha = 1,\cdots,k\,,
\end{equation}
where the $n$'s denote sets of $k$-ASEP occupation numbers, 
cf. Fig.\,\ref{processes}(b). But in view of the above mapping, each of these former 
corresponds to  a set of $\{\nu_1, \cdots, \nu_N\}$ occupations in the effective ASEP, 
so in particular it must hold that
\begin{equation}
\label{identity}
\sum_{\alpha = 1}^k\,\sum_{\;i \in \Lambda_{\alpha}} \left(1 - n_{i-k}\right)\;
n_{i-k+1} \cdots\,n_i = \sum_{j=1}^N\, \left( 1 - \nu_j \right)\,\nu_{j+1}\,.
\end{equation}
Here, the left hand side just counts the number of feasible movements in a given 
$k$-ASEP configuration which in turn must coincide with those counted by the right 
hand terms in the equivalent ASEP state. At this point it is worth mentioning that despite 
that for PBC all $k$-ASEP configurations are equally likely, the correlators involved in 
the sublattice currents  (\ref{currents}) are not factorizable \cite{Barma2, Barma3}. 
However, since for large $N$ the ASEP correlators do decouple under PBC \cite{Ligget}, 
it is then clear that as a result of Eq.\,\eqref{identity} the net sum of these currents 
amounts to
\begin{equation}
\label{sum-J}
\sum_{\alpha = 1}^k J_{\alpha} \simeq \frac{k N}{L}\;\rho_{_{_{\!\!ASEP}}} 
\left( 1- \rho_{_{_{\!\!ASEP}}}\right)\,.
\end{equation}
So, when it comes to the growth rates of the interface representation, from 
Fig.\,\ref{processes}(b) we can readily identify them with the contribution of all 
currents crossing a given height location, \emph{i.e.} ${\rm v}_k = \sum_{\alpha} 
J_{\alpha}$, each contribution here being associated with probabilities of mutually 
exclusive events wherein the height can grow. As for the density of ASEP particles in 
(\ref{sum-J}), further to Eq.\,\eqref{density} note that PBC also impose $\sum_n s_n 
\equiv 0$\, (\emph{i.e.} $h_L - h_1 = \pm 1$), for which the vacancy numbers there 
involved are constrained to add up to $L/2$ \cite{note2}. Therefore, we are left with a 
chain of $N = L/2 + (L-{\cal L})/k$ ASEP sites and $\rho_{_{\!ASEP}} ^{-1} = 1+ 
k/[\,2\,(1-{\cal L}/ L)\,]$, so the growth velocity (\ref{sum-J}) simply reduces to
\begin{equation}
\label{velocity}
{\rm v}_k ({\cal L}) = \left( \frac{\,2\,}{k} + \frac{1\,}{1 - {\cal L}/L\,}\right)^{\!\!-1}.
\end{equation}
As expected, so long as the $k$-ASEP dynamics is not fully jammed, \emph{i.e.} 
${\cal L}<L$, the interface can grow with finite rates, in turn being independent of the 
vacancy ordering in the string or sector considered.

{\it Exponential growth of invariant sectors}.--- 
From the above discussion it follows that the periodicity of these interfaces constrains 
each string to include $L/2$ 0's, \emph{i.e.} $L/2$ charcaters $A_i \ne A_k$, and 
${\cal L}-L/2$\, 1's. Moreover, the IS lengths $\mathcal{L}$ are restricted to belong to 
the set ${\cal S}_L^{^{(k)}}= \left\{ L, L-k, L-2k, \,\cdots,L/2 \right\}$, since an integer 
number of $k$-mers should be required to complete the total length $L$. More 
specifically, in terms of the vacancy numbers this reads
\begin{equation}
\label{constraints}
\sum_{i=0}^{k-1}\,N_i = L/2,\;\;\; \sum_{i=1}^{k-1} (i+1) \,N_i \in
\mathcal{S}_L^{(k)}\,,
\vspace{-0.1cm}
\end{equation}
and thereby the total number of irreducible sequences can be expressed as 
\begin{equation}\label{IScounting}
\mathcal{I}_k(L) = \sum_{\{N_i\}}\,\!\!^{^\prime} M(\{N_i\})\,.
\end{equation}
Here, the primed sum is a mnemonic device reminding us that the sum only goes over 
the sets $\{N_i\}$ complying with \eqref{constraints}, and $M(\{N_i\})$ denotes their
``multiplicities'', \emph{i.e.} the number of different orderings of the irreducible 
characters of the string. For PBC these orderings are counted up to cyclic permutations
of those characters, so that they are given by the ``circular multinomial coefficient''
\cite{Riordan}
\begin{align}
M(\{N_i\}) &= \frac{1}{N_0 + N_1 + \cdots + N_{k-1}} \sum_{d | \gcd(\{N_i\})} 
\varphi(d) \binom{\frac{N_0+\dots+N_{k-1}}{d}}{\tfrac{N_0}{d} \quad \cdots \quad 
\tfrac{N_{k-1}}{d}}\nonumber\\
&= \frac{2}{L} \sum_{d | \gcd(\{N_i\})} \varphi(d) \binom{\tfrac{L}{2d}}{\tfrac{N_0}
{d} \quad \cdots \quad \tfrac{N_{k-1}}{d}}\,,
\end{align}
where $\varphi(n)$ is the Euler's totient function \cite{Abramowitz}, and in the second 
line we used the first constraint of Eq.\,\eqref{constraints}.

For $k = 2$, we can perform the sum in \eqref{IScounting} to see that
\begin{equation}
\mathcal{I}_2(L) = \frac{1}{L}\,\sum_{d \big|\tfrac{L}{2}} 
\varphi\left(\tfrac{L}{d}\right) 2^d\,,
\end{equation}
so that in the $L \to \infty$ limit the sum over the divisors of $L/2$ is always dominated 
by the term $d = L/2$ and the number of invariant sectors grows as $2^{L/2} L^{-1}$.
For $k\geq3$ it is harder to do an exact calculation, but  we  can find  numerically the 
rates at which these sectors grow. These are listed in Table~\ref{periodic_tk} where it is 
clear that $\mathcal{I}_k(L) \simeq 2^L$ as $k\to\infty$, as was to be expected.
\vskip -0.15cm
\begin{table}[h!]
\centering
\begin{tabular}{|c|c|c|c|c|c|c|}
\hline
\hline
$k$ & 2 & 3 & 4 & 5 & 6 & 7\\
\hline
$\mathcal{I}_k^{1\!/\!L}\!(L) \propto$ & 1.41421 & 1.73205 & 1.8999 & 1.958 
&  1.981 & 1.991\\
\hline
\hline
\end{tabular}
\caption{\label{periodic_tk} Growth rates for the total number of sectors 
of motion [\,Eq.\,\eqref{IScounting}\,] under PBC.}
\end{table}
\vskip 0.1cm

We can also obtain a lower bound $\mathcal{I}^*_k(L)$ for $\mathcal{I}_k(L)$ by 
noting that
\begin{equation}
M(\{N_i\}) \geq M^*(\{N_i\}) \equiv \frac{1}{N_0 + N_1 + \cdots + N_{k-1}} 
\binom{N_0 + N_1 + \cdots + N_{k-1}} {N_0 \quad N_1 \quad \cdots \quad N_{k-1}}\,,
\end{equation}
so that using \eqref{constraints} we have
\begin{equation}
\mathcal{I}_k(L) \geq \mathcal{I}^*_k(L) \equiv \frac{2}{L} \sum_{\{N_i\}}\,\!
\!^{^\prime} \binom{L/2}{N_0 \quad \cdots \quad N_{k-1}}\,.
\end{equation}
For $k = 2$ this bound gives
\begin{equation}
\mathcal{I}_2^*(L) = \frac{2}{L}\,\sum_{i=0}^{L/4} \binom{L/2}{2i} = 
\frac{2^{L/2}}{L}\,,
\end{equation}
so it correctly captures the rate we had already found. For $k = 3$ one may show 
that $\mathcal{I}_3^*(L)$ satisfies the recursion
\begin{equation}
\frac{L}{2}\,\mathcal{I}_3^*(L) = \left(\,\tfrac52L - 4\right)\, \mathcal{I}_3^*(L-2) - 3 
\left(\tfrac12L-2\right) \left[\,\mathcal{I}_3^*(L-4) + 3\,\mathcal{I}_3^*(L-6)\,\right]\,,
\end{equation}
so that for large $L$ the total number of invariant sectors is at least
\begin{equation}
\mathcal{I}_3^*(L) \simeq \frac{3^{\tfrac{L}{2}-1}}{2} \left(1 + 3\sqrt{\tfrac{3}
{2\pi L}}\right)\,.
\end{equation}
Once more, the bound given by $\mathcal{I}_3^*(L)$ seems to be tight in the 
$L\to\infty$ limit, and we can check numerically that this is also the case for $k > 3$.

{\it Roughening exponents}.--- 
Armed with the effective ASEP correspondence referred to earlier on, we extensively 
sampled the stationary configurations of both dimer and trimer interfaces in some 
periodic IS sectors. These are specified in Table~\ref{tab2} along with their growth 
velocities [\,Eq.\,\eqref{velocity}\,], and sublattice densities (arising from simple 
stoichiometric considerations). Each state was prepared by random deposition of 
$N_k = (L- {\cal L})/k$ `monomers' on a ring of $L/2+ N_k$ effective sites 
[\,cf.\,Eq.\,\eqref{constraints}\,] which, depending on their  locations and occupancies, 
were then transformed to $k$-ASEP configurations according to the mapping discussed 
before. This enabled us to implement a sampling algorithm with a number of operations 
bounded as ${\cal O}[\,{\rm N}_s (L/2+N_k)\,]$, while using a number of samples 
${\rm N}_s$ such that  ${\rm N}_s \sim 10^{11}/L$ and substrates sizes of up to 
$10^6$ locations, thus significantly reducing the scatter of averaged data.

\begin{table}[htbp]
\begin{center}
\begin{tabular} { c c c c c c} \hline \hline 
$k$ & \hskip 0.5cm IS sector  &   \hskip 0.4cm Density
& \hskip 0.3cm  Growth rate & \hskip 0.8cm $\langle\, W^2 \rangle /L$
 & \hskip 0.75cm $\langle\,{\rm h_m}\,\rangle /L^{1/2}$  \vspace{0.1cm} \\  \hline 
 \vspace{-0.2cm} \\

&  \hskip 0.5cm
$[A_{_0}]^{L/2}$  &   \hskip 0.4cm 1/2 &  \hskip 0.3cm 1/3 & \hskip 0.8cm  0.125(1)
& \hskip 0.65cm 0.766(2) \vspace{0.1cm}\\

dimers  &  \hskip 0.5cm
$[A_{_1} A_{_0}^2]^{L/6}$ & \hskip 0.4cm $\begin{cases}\rho_1= 2/3
\vspace{0.125cm} \cr \rho_2 = 1/3 \end{cases}$ &  \hskip 0.3cm 1/4 & 
\hskip 0.8cm  0.074(1) & \hskip 0.65cm 0.583(2) \vspace{0.1cm}\\

&  \hskip 0.5cm
$[A_{_1} A_{_0}]^{L/4}$  &  \hskip 0.4cm  1/2  &  \hskip 0.3cm  1/5 &
\hskip 0.8cm  0.052(1) &  \hskip 0.65cm 0.494(2) \vspace{0.3cm}\\
\hline \\
\vspace{-0.7cm}\\

&  \hskip 0.5cm
$[A_{_0}]^{L/2}$  &   \hskip 0.4cm 1/2 &  \hskip 0.3cm 3/8 & \hskip 0.8cm  0.166(1)
& \hskip 0.65cm 0.881(1) \vspace{0.1cm}\\

trimers &  \hskip 0.5cm
$[A_{_2} A_{_0}^3]^{L/8}$  &  \hskip 0.5cm $\begin{cases} \rho_1 = \rho_2 = 5/8 
\vspace{0.125cm} \cr \rho_3 = 1/4 \end{cases}$ & \hskip 0.3cm  3/14 & \hskip 0.8cm  
0.073(1) & \hskip 0.65cm 0.584(1) \vspace{0.1cm}\\

 &  \hskip 0.5cm
$[A_{_2} A_{_1} A_{_0}^3]^{L/10}$  &  \hskip 0.4cm 1/2  &  \hskip 0.3cm 3/17 
& \hskip 0.8cm  0.056(2) & \hskip 0.65cm 0.515(2)  \vspace{0.25cm}\\
\hline \hline
\end{tabular}
\end{center}
\caption{Sublattice densities, growth velocities, and amplitudes of average widths and 
maximal heights for dimers and trimers in the irreducible strings studied in the main 
panels of Figs.\,\ref{exponents} and \ref{distrib}. Sectors of motion are formed by 
concatenating the string characters of Eq.\,\eqref{characters}, e.g.  $[A_1 
A_0^2]^{L/6}$ just repeats $[(10)(0)(0)] \; L/6$ times (so, ${\cal L}/L = 2/3$), etc.}
\label{tab2}
\end{table}

\begin{figure}[htbp]
\hskip -12.2cm
\includegraphics[width=0.34\textwidth]{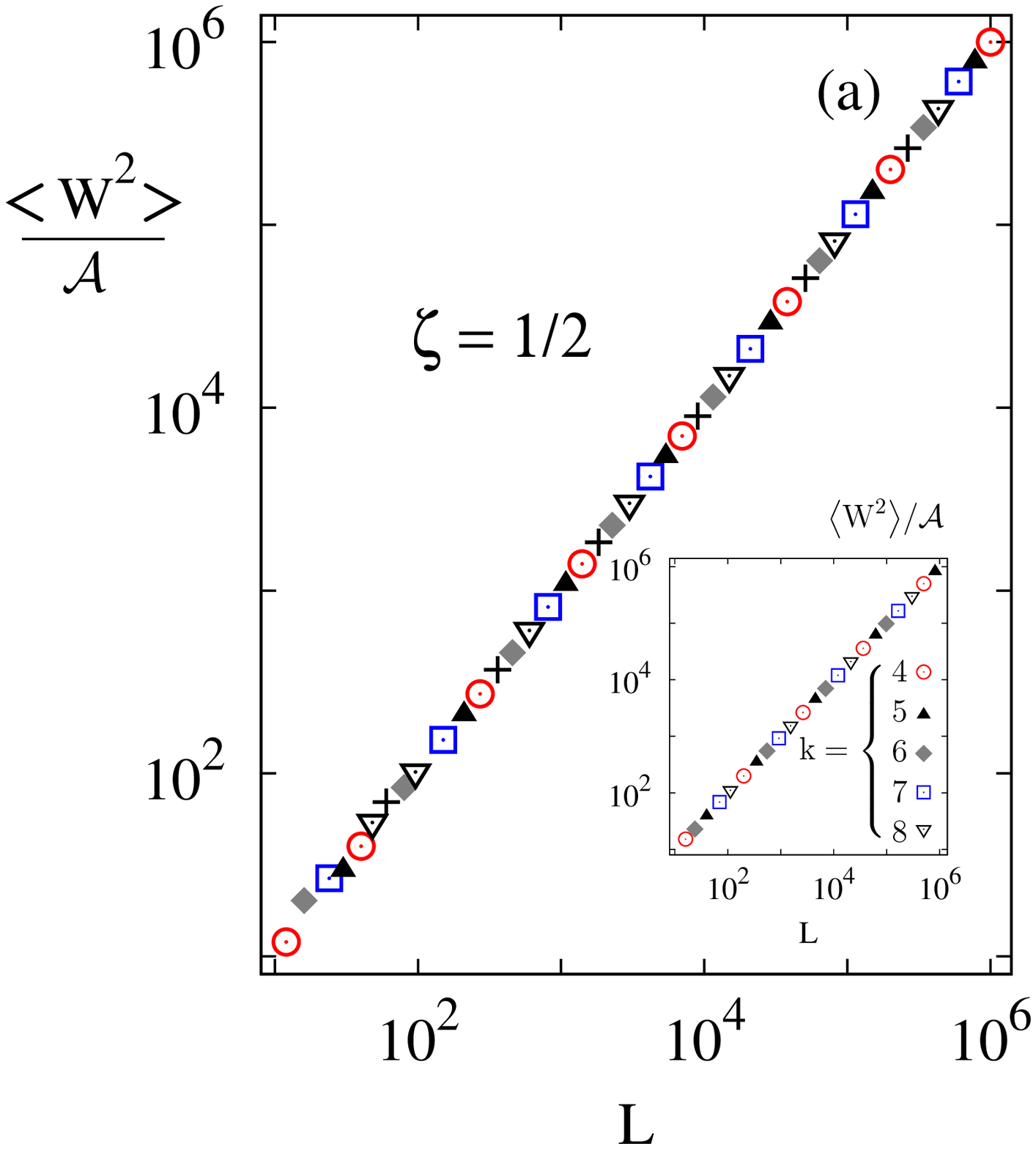}
\vskip -6.8cm
\hskip 0.1cm
\includegraphics[width=0.33\textwidth]{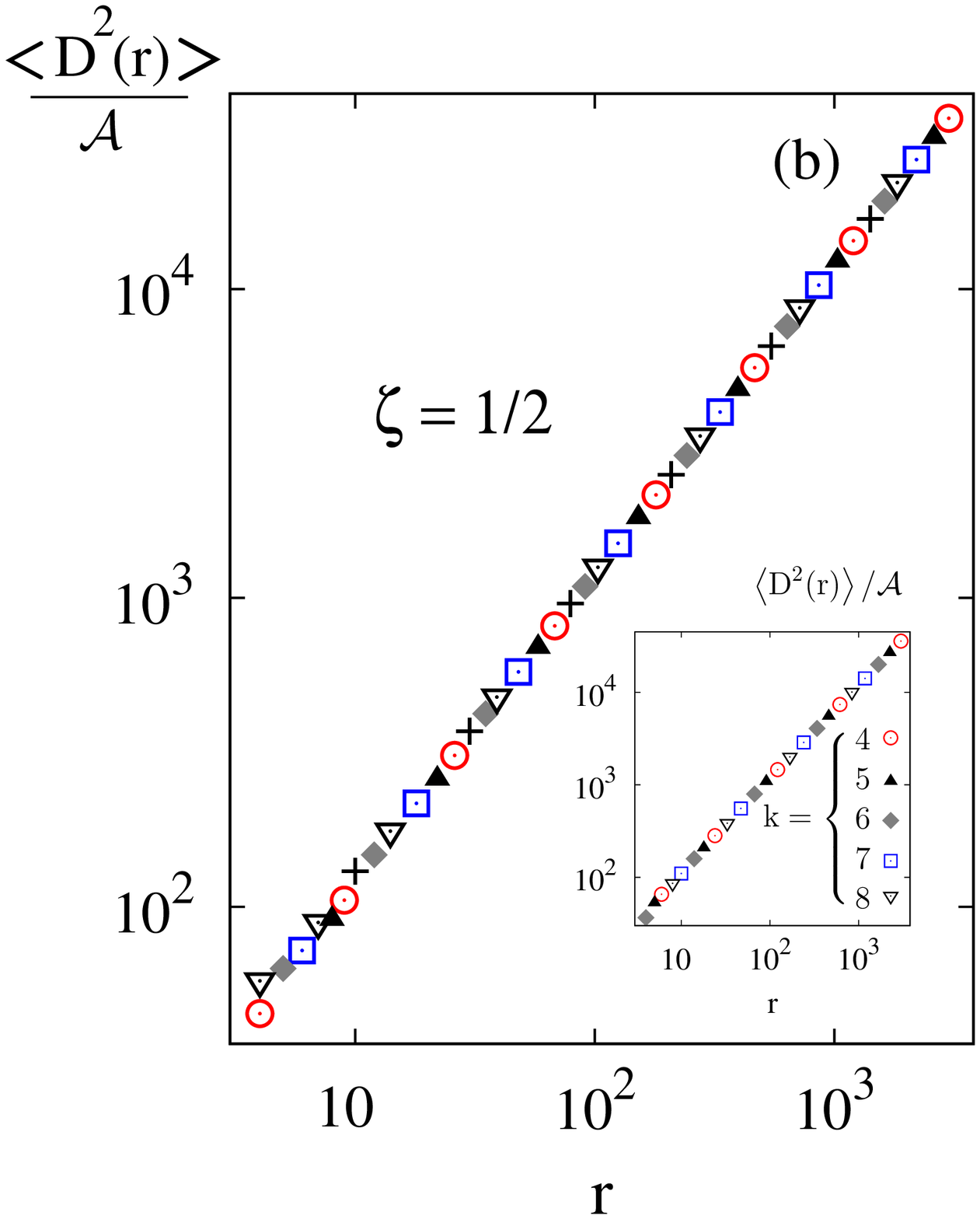}
\vskip -6.84cm
\hskip 11.78cm
\includegraphics[width=0.335\textwidth]{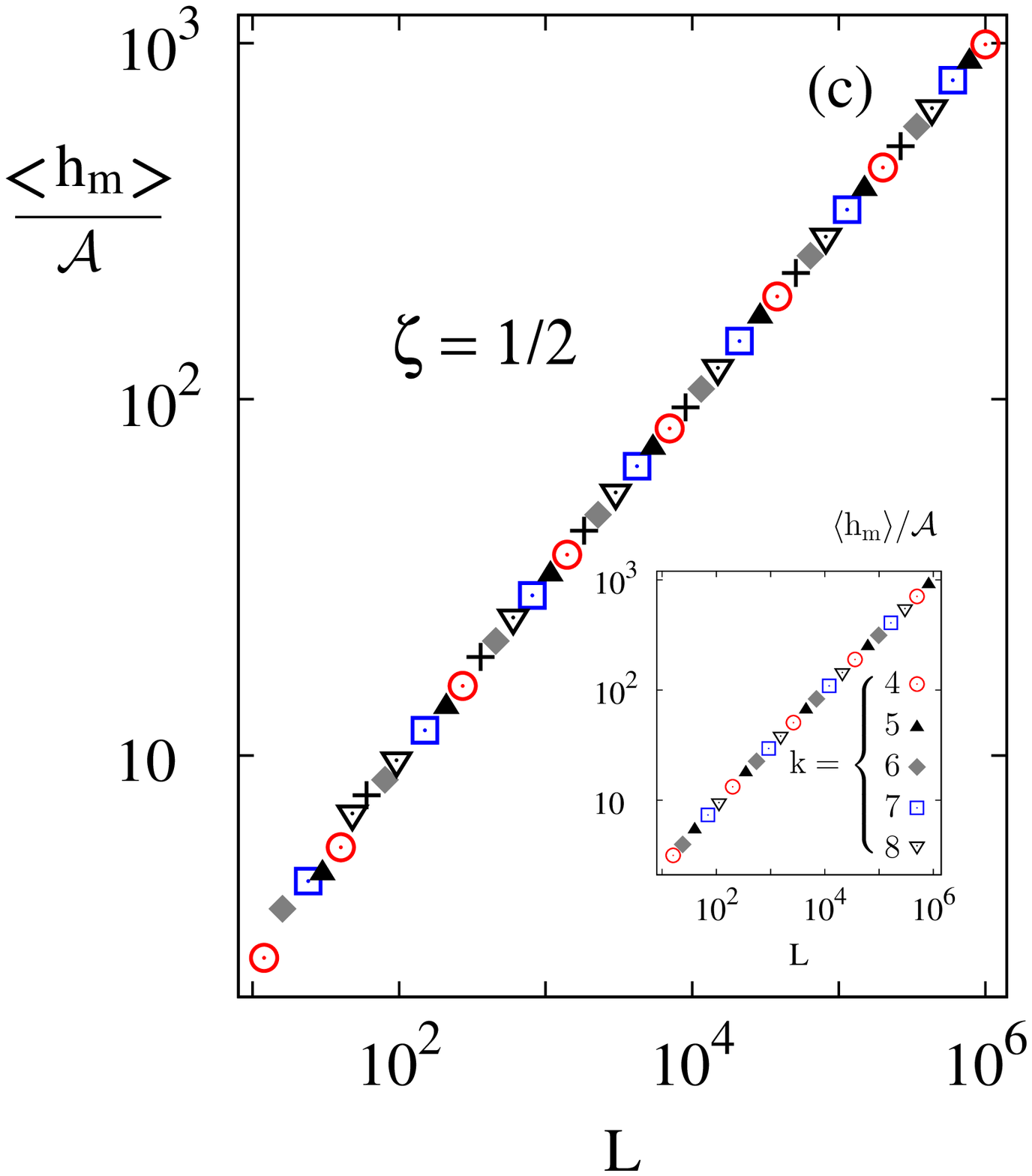}
\caption{(a) Finite-size growth of average widths, (b) height difference correlation 
functions ($L = 4.2 \times10^5$), and (c) growth of average maximum heights 
(measured with respect to spatially averaged ones), for various string sectors. In 
main panels, circles, squares, and rhomboids stand respectively for the first three 
sectors of Table~\ref{tab2} ($k=2$), whereas its following ones correspond to triangles, 
downwards triangles, and plus signs ($k=3$). Insets denote in turn cases of null string 
sectors with $k \ge 4$. All data in main panels (insets) were normalized by $\cal 
A$-\,amplitudes given in the rightmost columns of Table~\ref{tab2} (Table~\ref{tab3}), 
and are consistent with a common roughening exponent $\zeta =1/2$.}
\label{exponents}
\end{figure}

In Fig.\,\ref{exponents}(a) we exhibit the growth of the stationary widths (\ref{width}) 
spread over several decades of substrate lengths. As anticipated in the introductory 
paragraphs, despite the correlated movements and partitioning of the interface paths, 
all cases evidence the appearance of diffusive roughening exponents typical of monomer 
growing interfaces either in the KPZ or EW classes \cite{Meakin, KPZ,Edwards}. 
For display purposes, here the width of each dynamic sector was rescaled by the 
corresponding amplitudes of Table~\ref{tab2}, in turn decreasing with their growth 
velocities (as they should). 

Alongside Eq.\,\eqref{width} we also examined the stationary height difference 
correlation functions for which a similar scaling behavior involving the same 
roughening exponent is also expected to hold at distances $\vert r \vert \ll L$, 
that is \cite{Meakin,Family}
\begin{equation}
\label{corr}
\langle \,D_{_{\!\!L}}^2 (r) \,\rangle = \frac{1}{L} \,\sum_n \left\langle \big[\,h_{n+r} - 
\,h_n  \,\big]^2 \right\rangle \propto \vert r \vert ^{2 \zeta}.
\end{equation}
In fact, this is corroborated in Fig.\,\ref{exponents}(b) where these correlations turn out 
to scale linearly with the height separation for all IS sectors considered. As before, the 
results were made to collapse by the normalization amplitudes used for the average
widths of Fig.\,\ref{exponents}(a), as would be expected on the basis of the identity 
$\lim_{_{_{\!\!\!\!\!\!\!\!\!\!\!r \to \infty}}} \lim_{_{_{\!\!\!\!\!\!\!\!\!\!\!L \to \infty}}}\! 
\langle D_{_{\!\!L}}^2 (r) \rangle = \lim_{_{_{\!\!\!\!\!\!\!\!\!\!\!L \to \infty}}}\!\!2\, 
\langle W_{_{\!\!L}}^2\rangle$.

Another stationary quantity of interest whereby the roughening exponent can also 
be tested concerns the average maximal height $\langle{\rm h_m}\rangle_{\!_L}$ 
measured with respect to the spatially averaged height of each interface realization, 
namely
\begin{equation}
\label{hmax}
\left \langle\, {\rm h_m} \right \rangle_{\!_L} = \left \langle\, \max \left\{ h_1 -
\bar h,\,\cdots\,, h_L - \bar h \right\}\,\right \rangle \propto L^{\zeta}\,,
\end{equation}
thus capturing possible extreme fluctuations that neither the average width nor the 
height difference correlations are able to measure. The scaling of this quantity along with 
its stationary probability distribution (see below), have been investigated numerically 
\cite{Shapir} in discrete 1D growth models belonging to the EW class, as well as 
analytically \cite{Majumdar} applying path integral methods to both 1D EW and KPZ 
equations. In agreement with those studies, here also the correlated and partitioned 
paths described by our $k$-mer interfaces recover the diffusive scaling of $\langle{\rm 
h_m} \rangle_{\!_L}$ with the substrate size in all IS sectors of Table~\ref{tab2}. This 
is shown in Fig.\,\ref{exponents}(c) for a wide range of $L$ sizes after normalizing the 
data by the amplitudes of each sector. As might be presumed, these latter still decrease 
with their growth velocities and in all cases are quite larger than the corresponding width 
amplitudes (cf.\,Table~\ref{tab2}).

To complement the diffusive picture discussed so far, we also estimated the roughening 
exponents of Eqs.\,\eqref{width}, \eqref{corr}, and \eqref{hmax} in nonreconstituting
$[A_0]^{L/2}$ sectors of interfaces virtually grown out of several other $k$-mer values. 
This is displayed in the insets of Figs.\,\ref{exponents}(a), \ref{exponents}(b), and 
\ref{exponents}(c) where, just as in main panels, a $1/2$ scaling exponent can also be 
read off from their slopes. The normalizing amplitudes that produce the data collapse are 
quoted in Table~\ref{tab3}, and in parallel with the growth velocities $k / [2 (k+1)]$ 
these come out increasing monotonically with $k$, as was to be expected.

\vskip 0.25cm
\begin{table}[htbp]
\begin{center}
\begin{tabular} {c  c  c}
\hline \hline
\hskip 0.1cm
$k$ & \hskip 0.6cm  $\langle\, W^2 \rangle /L$  &  \hskip 0.5cm $\langle\,{\rm h_m}\,
\rangle /L^{1/2}$ \vspace{0.1cm} \\ 
\hline 
\vspace{-0.35cm} \\
4 &  \hskip 0.6cm 0.208(1)  & \hskip 0.132cm 0.987(1)
\\
5 & \hskip 0.6cm  0.250(1) & \hskip 0.1cm 1.083(2)
\\
6 &  \hskip 0.6cm 0.292(1)  & \hskip 0.1cm 1.162(5)
\\
7 & \hskip 0.6cm  0.334(1)   & \hskip 0.1cm 1.245(9)
\\
8 & \hskip 0.6cm  0.375(1)   & \hskip 0.1cm \;1.322(4)
\vspace{0.1cm}
\\
\hline \hline
\end{tabular}
\end{center}
\caption{Amplitudes of average widths and maximal heights for the null string sectors 
considered in the insets of Fig.\,\ref{exponents}.}
\label{tab3}
\end{table}

{\it Scaling distributions}.--- 
Turning to a more detailed level of description, next we focus our attention on the 
probabilities $P \left( {\rm h_m} \right), P \left( w^2 \right)$ of stationary realizations 
of both widths and maximal heights. Since their averages diverge in the thermodynamic 
limit, it has been argued on general grounds \cite{Zia, Shapir, Majumdar, Schehr, Racz, 
Parisi} that for large substrate sizes these probability distributions should scale as
\begin{equation}
P_{\!\!_L}\left( w^2 \right) \simeq \frac{1}{\langle\, W_{\!\!_L}^2 \,\rangle}\,
\Phi \left( \frac{w^2}{\langle\, W_{\!\!_L}^2 \,\rangle}  \right)\,,\;\;\;
P_{\!\!_L}\left( {\rm h_m} \right) \simeq \frac{1}{\langle\, {\rm h_m} \,\rangle_{\!_L}}\,
F \left( \frac{{\rm h_m}}{\langle\, {\rm h_m} \,\rangle_{\!_L}} \right)\,,
\end{equation}
where $\Phi (x)$ and $F(x)$ are characteristic scaling functions of a variety of 
solid-on-solid growth models \cite{Zia, Schehr}, although their dependence on 
boundary conditions is also a relevant issue \cite{Majumdar, Schehr, Halpin-Healy}. 

\begin{figure}[htbp]
\vskip 0.5cm
\hskip -8cm
\includegraphics[width=0.38\textwidth]{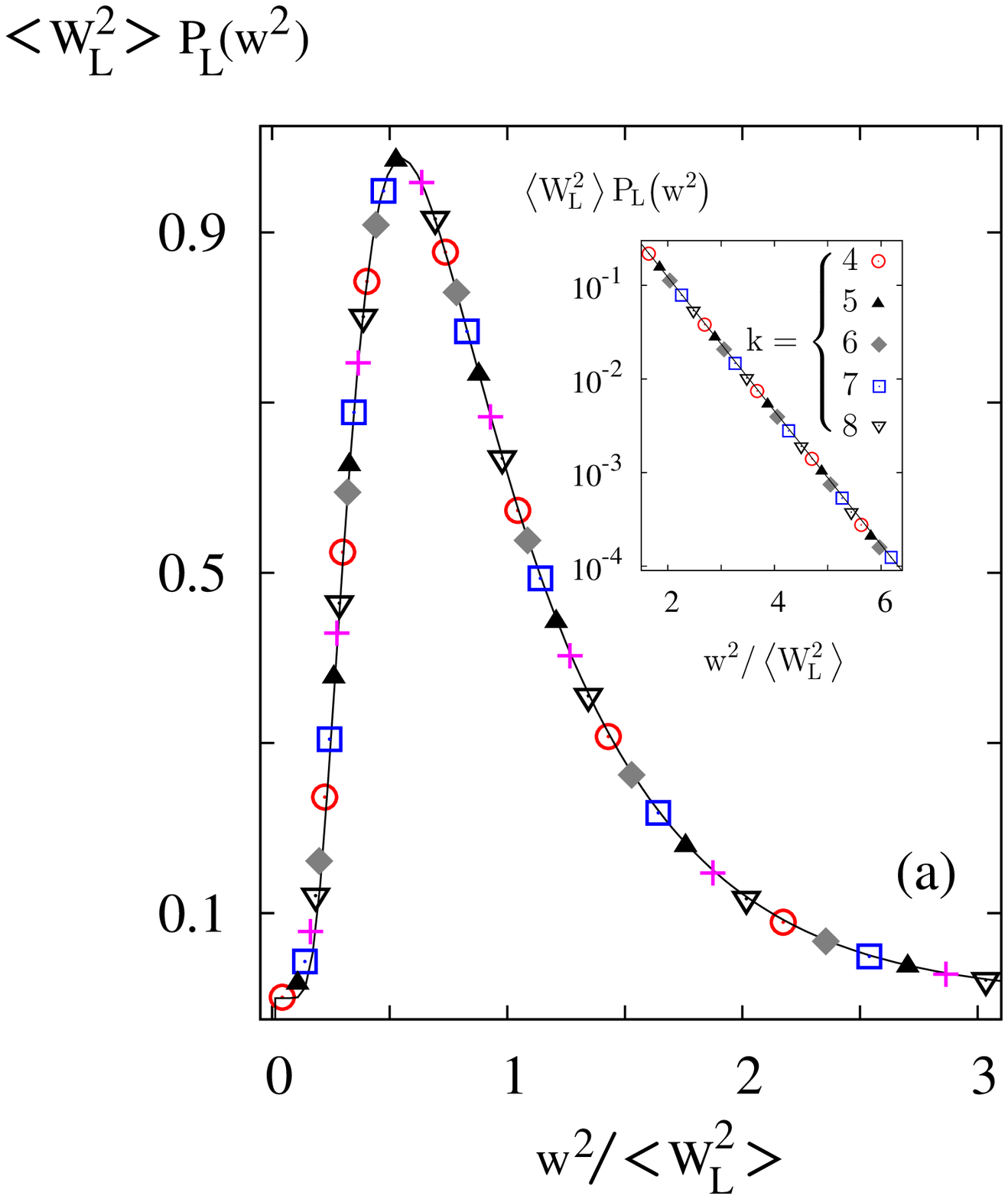}
\vskip -7.25cm
\hskip 6cm
\includegraphics[width=0.38\textwidth]{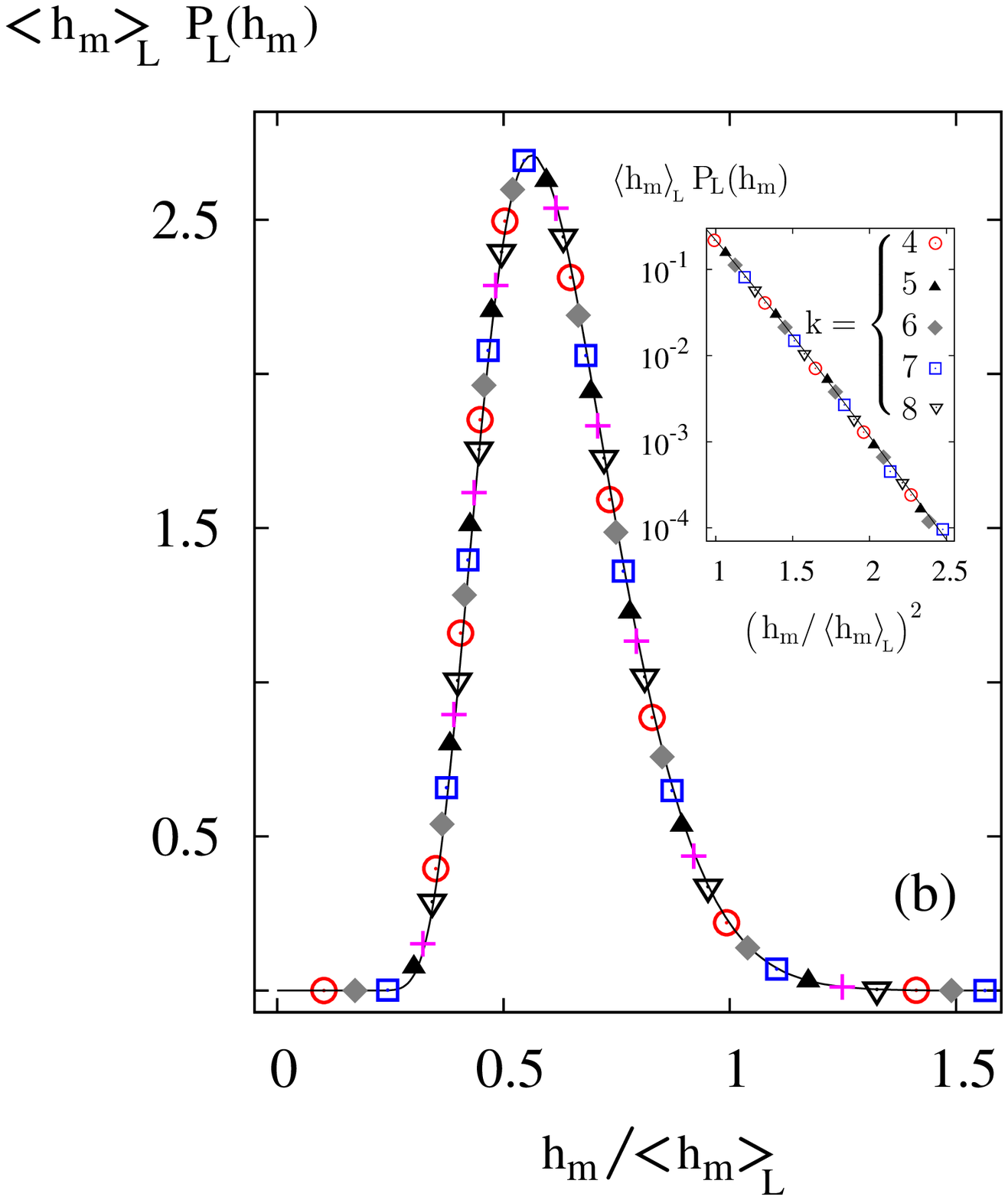}
\caption{Scaling of (a) width distributions, and (b) maximal height distributions for 
the $k$-mer sectors of Table~\ref{tab2} using the symbols of Fig.\,\ref{exponents}. 
Here, these stand for sizes $L=3 \times 10^4$ (circles), $6\times 10^4$ (rhomboids), 
$10^5$  (squares), $2.4 \times 10^4$ (triangles), $4.8 \times 10^4$ (plus signs),
and $9 \times 10^4$ (downwards triangles). For comparison, in (a) and (b) solid 
lines correspond respectively to the exact scaling functions referred to in Eq.\,\eqref{phi} 
(Ref.\,\cite{Zia}), and Eq.\,\eqref{F} (Ref.\,\cite{Majumdar}). The insets exhibit the tails 
of those scaling distributions which also follow our data in the null string sectors of $k=4,
\,5,\,6,\,7,\,8$ with $L= (\,3.6,\,4.2,\,3.3,\,2.8,\,4\,) \times\!10^4$ respectively.} 
\label{distrib}
\end{figure}

In particular under PBC, where the $k$-ASEP mapping has so far been applied, these 
scaling functions were evaluated exactly in 1D Brownian interfaces, thus enabling us 
to go a step further in the characterization of our $k$-mer models. This we do in 
Figs.\,\ref{distrib}(a) and \ref{distrib}(b) where the scaled probability distributions of 
$w^2$ and ${\rm h_m}$ in all IS sectors of Table~\ref{tab2} are compared with the 
analytical expressions of $\Phi$ and $F$ obtained respectively in Refs.\,\cite{Zia} and 
\cite{Majumdar}, namely
\begin{subequations}
\begin{eqnarray}
\label{phi}
\Phi (x) &=& \frac{\pi^2}{3}\,\sum_{n \ge 1}\,(-1)^{n-1}\,n^2\,
\exp \left(-\frac{\pi^2}{6} n^2 x \right)\,,\\
\label{F}
F (x) &=& \frac{2\,\sqrt 6}{x^{10/3}}\,\sum_{n \ge 1}\,b_n^{2/3}\,
\exp \left(- \frac{b_n}{x^2}\right)\, U \! \left(- \frac{5}{6}\,,\, 
\frac{4}{3}\,,\, \frac{b_n}{x^2} \,\right)\,.
\end{eqnarray}
\end{subequations}
Here $U (x_1, x_2, x_3)$ denotes the confluent hypergeometric function 
\cite{Abramowitz}, whereas $b_n \equiv 2 \left(\vert a_n \vert /3\right)^3$ involves the 
magnitudes of the Airy function zeros ($a_n$) on the negative real axis \cite{Majumdar,
Abramowitz}. Using substrates in the range of $10^4\,-\,10^5$ heights, the probability 
densities were reconstructed by means of the convolution of $10^7$ data points (in turn
derived from independent $k$-ASEP samples), with a Gaussian kernel whose bandwidth 
was determined by Silverman's method \cite{Silverman}.
In all sectors considered the data collapse is in excellent agreement with the scaling 
distributions (\ref{phi}) and (\ref{F}). Here, note that there are no parameters to 
fit these stationary functions and that no scaling properties neither for  $\langle\, 
W_{\!\!_L}^2 \,\rangle$ nor $\left \langle\, {\rm h_m} \right \rangle_{\!_L}$ have 
been used, the only approximation being the finite size of the substrates. The data 
collapse towards the tails of these distributions is also corroborated in the insets of 
Figs.\,\ref{distrib}(a) and \ref{distrib}(b) where other $k$-mer values are examined in 
nonreconstructing $[A_0]^{L/2}$ sectors. For large realizations of $w^2$ and ${\rm 
h}_m$ the resulting slopes of the semilogarithmic plots displayed there in fact coincide 
with those derived from the asymptotic behavior of $\Phi$ and $F$, decaying respectively 
as $\exp(- \frac{\pi^2}{6} x)$ and $\exp(- 6\,x^2)$ (cf.\,Refs.\,\cite{Zia,Majumdar}).

Further to periodic strings, we also considered disordered IS sectors obtained from 
the former by random permutations of their characters. It is worth mentioning that 
preliminary results also indicate that the above scaling distributions continue to stand 
as generic features of that disordered situation.

\begin{figure}[htbp]
\vskip 0.5cm
\hskip -8cm
\includegraphics[width=0.36\textwidth]{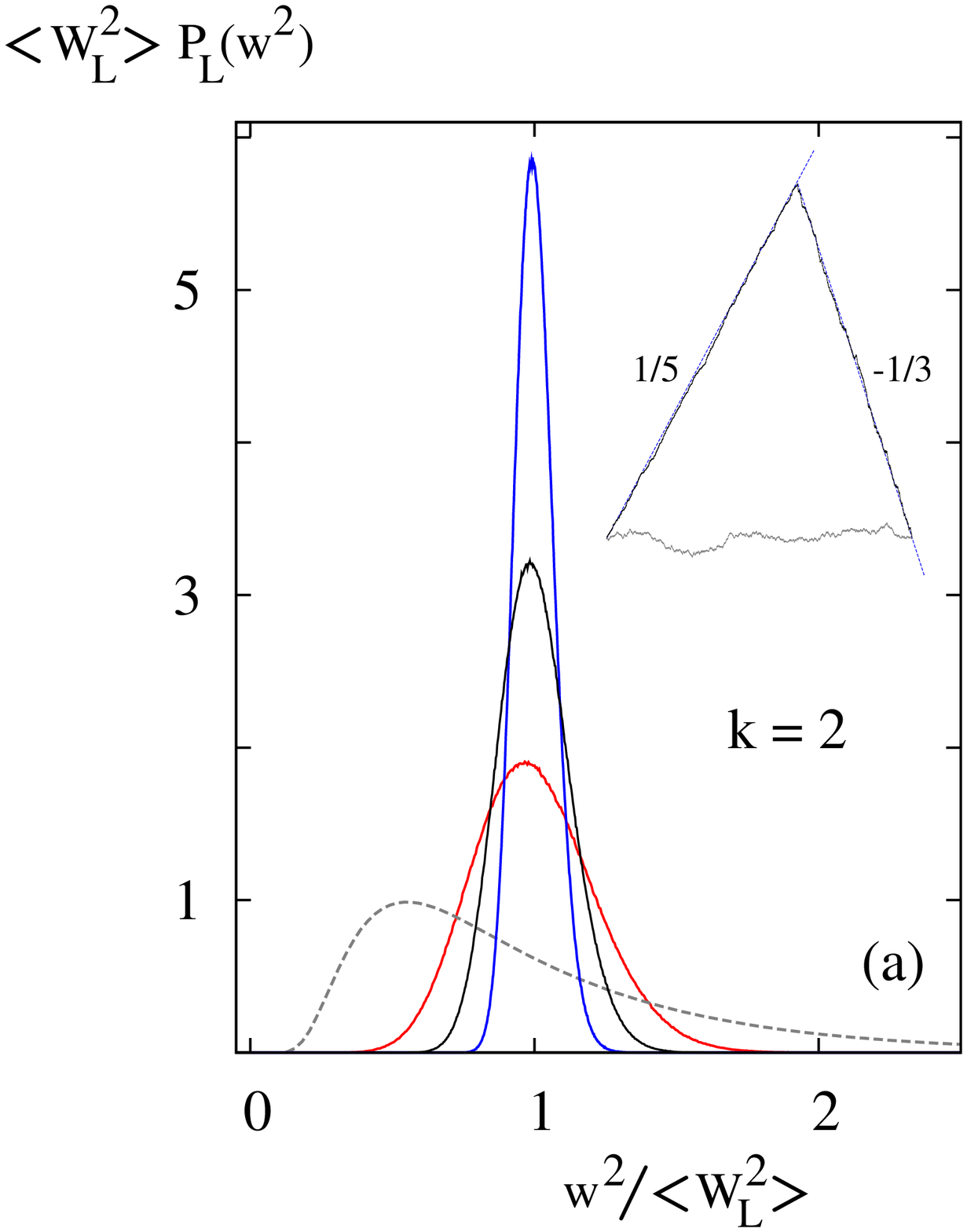}
\vskip -7.3cm
\hskip 6.5cm
\includegraphics[width=0.36\textwidth]{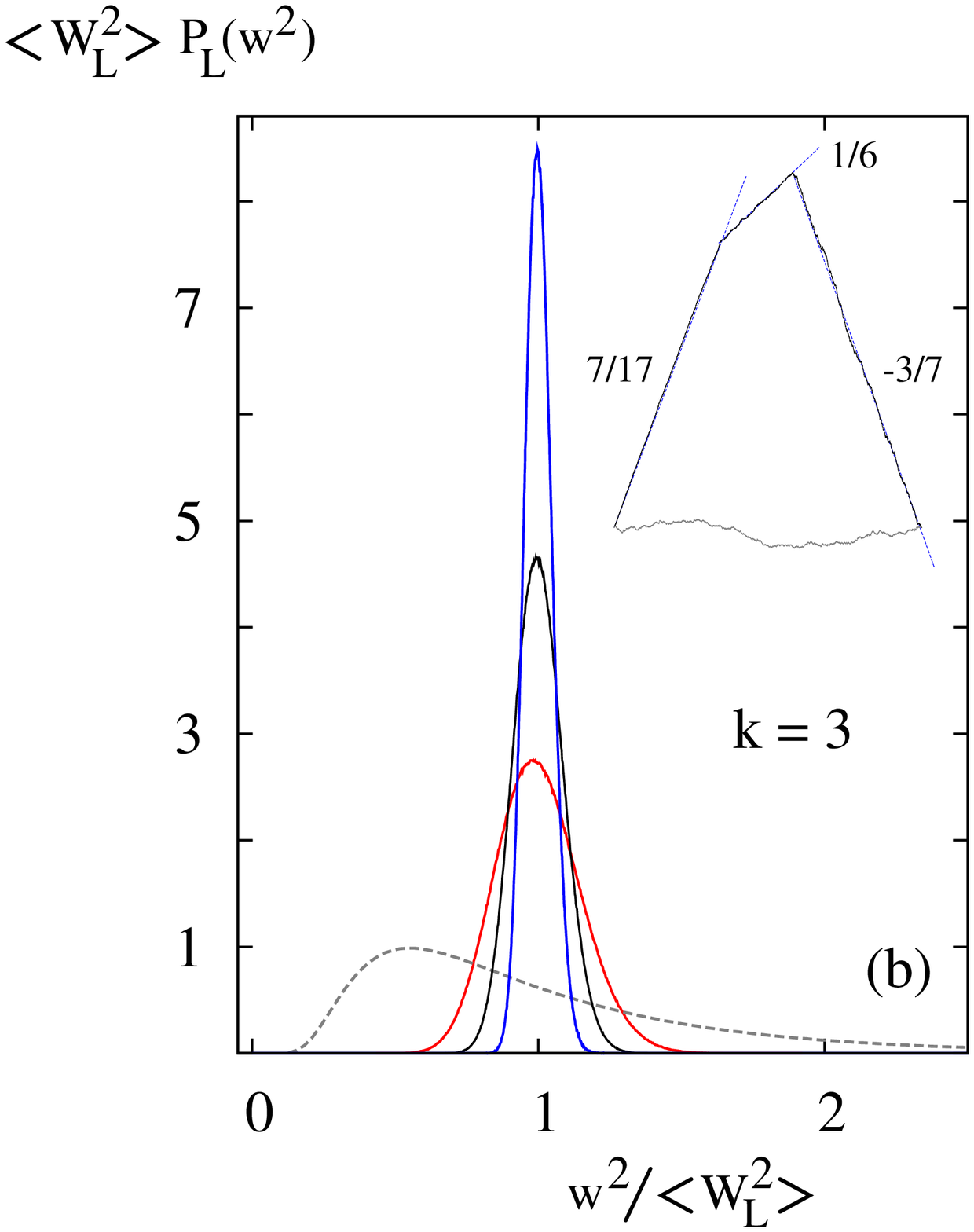}
\vskip 0.6cm
\caption{Width distributions of faceting sectors (a) $[A_1]^{L/4} [A_0]^{L/4}$ 
(${\cal L}/ L = 3/4$), and (b) $[A_2]^{L/10} [A_1]^{L/10} [A_0]^{3 L /10}$ 
(${\cal L}/ L = 4/5$) for $L = 10^4, 3 \times 10^3,$ and $10^3$ (topmost solid 
lines in downward direction). In contrast to roughening sectors (lowermost universal 
distribution), fluctuations around average widths become negligible as $L$ increases. 
Snapshots of the resulting interfaces ($L = 10^4$) are displayed by the insets. 
The slopes of their facets (indicated by dashed lines) are evaluated in the text. 
For comparison, snapshots of the roughening sectors $[A_1\,A_0]^{L/4}$, and 
$[A_2\, A_1 \,A_0^3\,]^{L/10}\!$ are also exhibited in (a) and (b) respectively.}
\label{facets}
\end{figure}

{\it Faceting}.---
Finally however, and in marked contrast with that robustness, let us comment on string 
sectors that include long concatenations of identical vacancy types, such as those 
considered in Fig.\,\ref{facets}. When the length of these domains becomes of the order 
of the substrate size, it turns out that the implicit assumption of a well-defined average 
orientation of the interface (parallel to the substrate) is no longer consistent. Instead, 
a faceted structure with large scale slopes emerges. This is illustrated by the snapshots 
shown in the insets of Figs.\,\ref{facets}(a) and \ref{facets}(b), each of their facets 
stemming from different character domains along their strings. Moreover, as suggested 
by the width distributions displayed in main panels, statistical fluctuations in these 
structures are progressively suppressed as $L$ increases.

In that latter respect we can assume a uniform density of effective ASEP particles for most 
interface realizations so as to readily estimate the slope of each facet. Thus, if there are 
$n$ characters in a given domain, clearly the average number of $k$-mers amid them 
should be $n (\frac{L-{\cal L}}{k})/ \sum_{i=0}^{k-1}\,N_i = \frac{2 n}{k} (1-{\cal L}
/L)$ [\,see PBC constraints of Eq.\,\eqref{constraints}\,]. Since each $A_j$ involves $j$ 
monomers and a vacancy, then the average length of such set ($A_j$-characters and 
$k$-mers combined) must comprehend $n (j+1) +2n(1-{\cal L}/L)$ sites of the substrate 
($n\le N_j$). Analogously, the average height difference along that set becomes $n (j-1) + 
2 n (1 - {\cal L}/L)$. Thereby, we are left with slopes $1-\left( \frac{ j+3}{2}-\frac{\cal L}
{L} \right)^{-1}$ that closely follow those arising from the $A_j$-domains considered in 
the strings of Fig.\,\ref{facets}. Note that these average slopes can vanish only in 
nonreconstituting $[A_0]^{L/2}$ sectors but, as seen above, in such cases the usual 
roughening behavior is restored.

To summarize, we have studied stationary aspects of 1D interfaces formed by deposition 
of extended particles within the context of a mapping to a process of driven and 
reconstituting $k$-mers \cite{Barma2,Barma3}. This enabled us to sample the steady 
state without having to explicitly evolve the system in time, and, as a result, a rich 
statistical analysis of both stationary width and maximal height distributions was 
attained at large substrate scales. For clarity of presentation the models were defined
as totally asymmetric, although extensions using partially asymmetric or even symmetric 
versions subject to PBC would make no difference to the stationary distributions. 

The notion of irreducible string played a key role in the understanding of the behavior 
of these interfaces as it encodes nonlocal conserved quantities that partition the growth 
dynamics into an exponential number of disjoint sectors of motion with specific growth 
velocities. Owing to the spatial extension of the deposited $k$-mers, the path phase 
space of these sectors actually corresponds to sets of correlated random walks. However, 
in view of the diffusive roughening exponents obtained for several IS sectors, these 
walks turn out to follow the typical root mean square displacement associated with the 
stationary roughness of the 1D KPZ and EW classes. Finally, at the more demanding level 
of width and maximal height probability distributions, all roughening sectors considered 
also reproduced the exact scaling functions \cite{Zia, Majumdar} of those universality 
classes. Whether these numerical findings could be explained theoretically remains an 
open issue which, in turn, should also account for the existence of faceting sectors. 

\section*{Acknowledgments}
M.D.G. acknowledges support from CONICET (PIP 2015-813) and ANPCyT (PICT 1724).
The work of F.I.S.M. was supported by IBS-R018-D2.\, F.I.S.M. would like to thank IFLP and 
UNLP for hospitality during the completion of this work.



\begin{thebibliography}{99}
\bibitem{note2} For simplicity, hereafter substrate sizes $L$ and string lengths $\cal L$ 
are chosen such that $L \propto 2 k$, and ${\cal L} \propto k$.
\end{thebibliography}
\end{document}